\documentclass[fleqn,usenatbib]{mnras}
\usepackage[T1]{fontenc}
\usepackage{ae,aecompl}
\usepackage{graphicx}	
\usepackage{amsmath}	
\usepackage{amssymb}	
\usepackage{subfigure}
\usepackage{threeparttable}
\usepackage{lineno}

\title[late-time transition of $H_{0}$]{Revealing the late-time transition of $H_{0}$: relieve the Hubble crisis}
\author[Jian-Ping, Hu et al.]{
J. P. Hu,$^{1}$
F. Y. Wang$^{1,2}$\thanks{E-mail: fayinwang@nju.edu.cn}
\\
$^{1}$School of Astronomy and Space Science, Nanjing University, Nanjing 210093, China\\
$^{2}$Key Laboratory of Modern Astronomy and Astrophysics (Nanjing University), Ministry of Education, Nanjing 210093, China
}

\date{Accepted XXX. Received YYY; in original form ZZZ}

\pubyear{2022}

\begin{document}
\label{firstpage}
\pagerange{\pageref{firstpage}--\pageref{lastpage}}
\maketitle

\begin{abstract}
The discrepancy between the value of the Hubble constant $H_0$ measured from the local distance ladder and from the cosmic microwave background is the most serious challenge to the standard $\Lambda$CDM model. Various models have been proposed to solve or relieve it, but no satisfactory solution has been given until now. Here, we report a late-time transition of $H_{0}$, i.e., $H_0$ changes from a low value to a high one from early to late cosmic time, by investigating the Hubble parameter $H(z)$ data based on the Gaussian process (GP) method. This finding effectively reduces the Hubble crisis by 70\%. Our results are also consistent with the descending trend of $H_0$ measured using time-delay cosmography of lensed quasars at 1$\sigma$ confidence level, and support the idea that the Hubble crisis arises from new physics beyond the standard $\Lambda$CDM model. In addition, in the $\Lambda$CDM model and $w$CDM model, there is no transition behavior of $H_{0}$.
\end{abstract}



\section{Introduction}\label{sec:intro}

While most astronomical observations \citep{2017MNRAS.470.2617A,2018ApJ...859..101S,2020A&A...641A...6P,2021A&A...647A..38B,2021MNRAS.504.2535I,2022MNRAS.513.5517C} are in agreement with the cosmological constant ($\rm \Lambda$) cold dark matter model ($\rm \Lambda$CDM), there is a severe tension of the Hubble constant, $H_{0}$, by assuming the $\Lambda$CDM model. There has been a 4.0-5.9$\sigma$ tension \citep{2019ApJ...882...34F,2019NatAs...3..891V,2020MNRAS.498.1420W} between the $H_{0}$ values inferred from the \emph{Planck} cosmic microwave background \citep[CMB;][]{2020A&A...641A...6P} data and the local distance ladder \citep{2019ApJ...876...85R,2022ApJ...934L...7R}. In order to solve or relieve the $H_{0}$ crisis, a lot of theoretical models have been proposed, which can be broadly divided into three categories, including early-time models, late-time models and modified-gravity models. Some recent reviews on the $H_{0}$ crisis can be found in \cite{2020NatRP...2...10R,2021CQGra..38o3001D,2021A&ARv..29....9S,2022PhR...984....1S}.

Recently, a possible trend between the lens redshift and the inferred of $H_{0}$ was found using time-delay cosmography of lensed quasars by the H0LiCOW collaboration \citep{2020MNRAS.498.1420W}. The statistical significance level is about 1.9$\sigma$. After that, \citet{2020A&A...639A.101M} added a new H0LiCOW lens (DES J0408-5354) which makes the tentative trend slightly reduced to 1.7$\sigma$. The TDCOSMO IV re-analysis including H0LiCOW data lowers the $H_{0}$ value and increases the error bar \citep{2020A&A...643A.165B}. So, the significance of the H0LiCOW $H_{0}$ trend may be lower. \citet{2020PhRvD.102j3525K} constrained the cosmological parameters in different redshift ranges ($z <$ 0.7) by binning the cosmological dataset comprising megamasers, cosmic chronometers (CC), type Ia supernovae (SNe Ia) and Baryon Acoustic Oscillation (BAO) according their redshifts. They also found a similar $H_0$ descending trend with low significance in different cosmological models \citep{2020PhRvD.102j3525K}. In the following, this binning method is referred to the partition method. The same $H_0$ descending trend has also been found from other observational data including SNe Ia \citep{2021ApJ...912..150D,2021arXiv211103055H,2022arXiv220310558C} and quasars\citep{2022arXiv220310558C}. If this trend is true, it will provide a potentially innovative solution for the Hubble crisis. However, no research verifies this trend basing on a model-independent method. There has been a lot literature \citep{2019JCAP...12..035K,2019MNRAS.483.4803L,2020A&A...643A.165B}, which alleviate the present Hubble crisis utilizing model independent analyses of $H(z)$ data.\\

\begin{figure}
	\centering
	\includegraphics[width=0.4\textwidth]{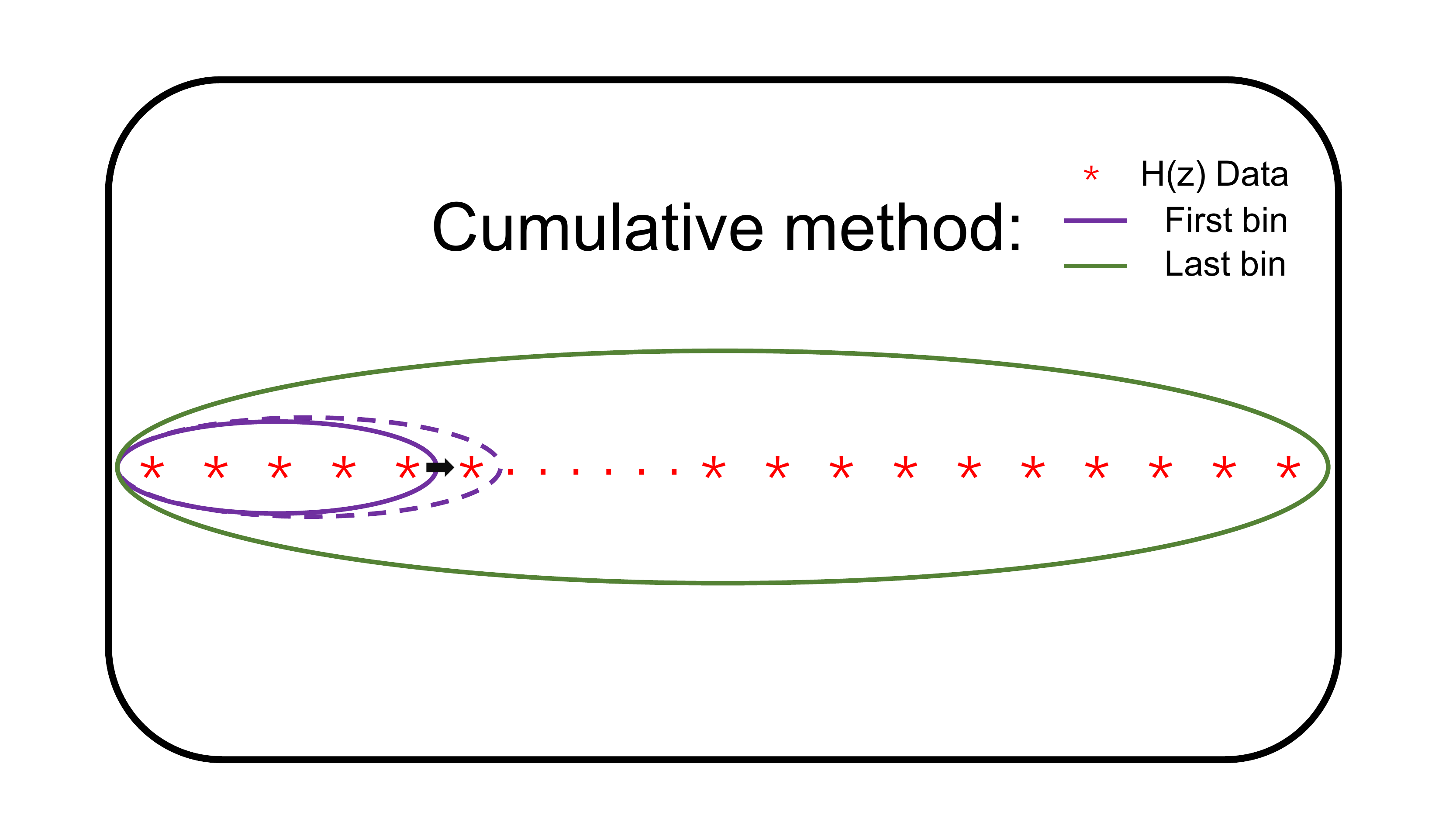}
	\caption{\label{Fig1} Schematic diagram for the cumulative binning method.}
\end{figure}

In this paper, we investigate the redshift-evolution of $H_{0}$ with 36 Hubble parameter $H(z)$ data \citep{2018ApJ...856....3Y} adopting the GP method \citep{2012arXiv1201.0490P}, which was widely used in cosmological researches \citep{2010PhRvL.105x1302H,2012MNRAS.425.1664B,2018JCAP...04..051G,2018JCAP...02..034M,2018ApJ...856....3Y,2019MNRAS.486L..46A,2019JCAP...12..035K,2019ApJ...886L..23L,2020ApJ...895L..29L,2021MNRAS.507..730H,2022ApJ...924...97W}. The use of GP method to derive $H_{0}$ avoids a assumption of cosmic model. Unlike previous works \citep{2020PhRvD.102j3525K,2021ApJ...912..150D}, we employ a different binning method, i.e., the next bin has one more high-redshift data than the previous bin, which is illustrated in Figure \ref{Fig1}.
We refer to this binning method as the cumulative method. The distinction between the partition and the cumulative methods will be discussed in more detail in the next section. If $H_{0}$ evolving with redshift is substantiated and consistent with other observations \citep{2019ApJ...876...85R,2020A&A...641A...6P,2020MNRAS.498.1420W}, it can be regarded as a potential solution to the $H_{0}$ crisis.

\section{Data and methods}

The Hubble parameter measurements are taken from \cite{2018ApJ...856....3Y}. There are 36 $H(z)$ data covering redshift range (0.07, 2.36). In this catalog, 31 $H(z)$ are derived by comparing relative ages of galaxies at different redshifts. The formula is given by \citep{2002ApJ...573...37J}
\begin{equation}
	H(z) = -\frac{1}{1+z} \frac{d z}{d t}.
	\label{Hz}
\end{equation}
Based on the measurements of the age difference, $\Delta t$, between two passively evolving galaxies that are separated by a small redshift interval $\Delta z$, the value of \emph{dz/dt} can be approximately replaced with $\Delta z$/$\Delta t$. The cosmic chronometric $H(z)$ error bars are dominated by systematic uncertainty, which has been extensively discussed \citep{2016JCAP...05..014M}. Three correlated measurements are from the radial BAO signal in galaxy distribution. Based on this method, the angular diameter distance $D_{A}$ and Hubble parameter $H(z)$ can be constrained \citep{2017MNRAS.470.2617A}. Here, we present the covariance matrix of the three galaxy distribution radial BAO $H(z)$ measurements \citep{2017MNRAS.470.2617A}

\begin{equation}\label{eq:matrix}
	\left(
	\begin{array}{ccc}
		3.65	&	1.78	&	0.93	\\
		1.78	&	3.65	&	2.20	\\
		0.93	&	2.20	&	4.45	\\
	\end{array}
	\right)
\end{equation}
and take it into account in our calculations. The last two $H(z)$ data are measured from the BAO signal in the Ly$\alpha$ forest distribution alone or cross-correlated with quasar observations \citep{2014JCAP...05..027F,2015AA...574A..59D}. It is worth noting that all $H(z)$ measurements do not depend on the choice of $H_{0}$. The $H(z)$ data is shown in Table \ref{tab:Hzdata}.
\linespread{1.00}
\begin{table}\footnotesize
	\caption{ $H(z)$ data (in units of $\textrm{km}~\textrm{s}^{-1} \textrm{Mpc}^{-1}$). \label{tab:Hzdata}}
	\centering
	\begin{tabular}{ccc}
		\hline\hline
		$z$  & $H(z)$  & Reference    \\
		\hline
		$0.07$    & $69.0\pm19.6$$^{a}$ & \citep{2014RAA....14.1221Z}\\
		$0.09$    & $69.0\pm12.0$$^{a}$ & \citep{2005PhRvD..71l3001S}\\
		$0.12$    & $68.6\pm26.2$$^{a}$ & \citep{2014RAA....14.1221Z}\\
		$0.17$    & $83.0\pm8.0$$^{a}$ &  \citep{2005PhRvD..71l3001S}\\
		$0.179$   & $75.0\pm4.0$$^{a}$ &  \citep{2012JCAP...08..006M}\\
		$0.199$   & $75.0\pm5.0$$^{a}$ &  \citep{2012JCAP...08..006M}\\
		$0.2$     & $72.9\pm29.6$$^{a}$ & \citep{2014RAA....14.1221Z}\\
		$0.27$    & $77.0\pm14.0$$^{a}$ & \citep{2005PhRvD..71l3001S}\\
		$0.28$    & $88.8\pm36.6$$^{a}$ & \citep{2014RAA....14.1221Z}\\
		$0.352$   & $83.0\pm14.0$$^{a}$ & \citep{2012JCAP...08..006M}\\
		$0.38$    & $81.9\pm1.9$$^{b}$ &  \citep{2017MNRAS.470.2617A}\\
		$0.3802$  & $83.0\pm13.5$$^{a}$ & \citep{2016JCAP...05..014M}\\
		$0.4$     & $95.0\pm17.0$$^{a}$ & \citep{2005PhRvD..71l3001S}\\
		$0.4004$  & $77.0\pm10.2$$^{a}$ & \citep{2016JCAP...05..014M}\\
		$0.4247$  & $87.1\pm11.2$$^{a}$ & \citep{2016JCAP...05..014M}\\
		$0.4497$ & $92.8\pm12.9$$^{a}$ &  \citep{2016JCAP...05..014M}\\
		$0.47$    & $89.0\pm50$$^{a}$  &  \citep{2017MNRAS.467.3239R}\\
		$0.4783$  & $80.9\pm9.0$$^{a}$ &  \citep{2016JCAP...05..014M}\\
		$0.48$    & $97.0\pm62.0$$^{a}$ &  \citep{2017MNRAS.467.3239R}\\
		$0.51$    & $90.8\pm1.9$$^{b}$ &   \citep{2017MNRAS.470.2617A}\\
		$0.593$   & $104.0\pm13.0$$^{a}$ & \citep{2012JCAP...08..006M}\\
		$0.61$    & $97.8\pm2.1$$^{b}$ &   \citep{2017MNRAS.470.2617A}\\
		$0.68$    & $92.0\pm8.0$$^{a}$  &  \citep{2012JCAP...08..006M}\\
		$0.781$   & $105.0\pm12.0$$^{a}$ & \citep{2012JCAP...08..006M}\\
		$0.875$   & $125.0\pm17.0$$^{a}$ & \citep{2012JCAP...08..006M}\\
		$0.88$    & $90.0\pm40.0$$^{a}$  & \citep{2017MNRAS.467.3239R}\\
		$0.9$     & $117.0\pm23.0$$^{a}$ & \citep{2005PhRvD..71l3001S}\\
		$1.037$   & $154.0\pm20.0$$^{a}$ & \citep{2012JCAP...08..006M}\\
		$1.3$     & $168.0\pm17.0$$^{a}$ & \citep{2005PhRvD..71l3001S}\\
		$1.363$   & $160.0\pm33.6$$^{a}$ & \citep{2015MNRAS.450L..16M}\\
		$1.43$    & $177.0\pm18.0$$^{a}$ & \citep{2005PhRvD..71l3001S}\\
		$1.53$    & $140.0\pm14.0$$^{a}$ & \citep{2005PhRvD..71l3001S}\\
		$1.75$    & $202.0\pm40.0$$^{a}$ & \citep{2005PhRvD..71l3001S}\\
		$1.965$   & $186.5\pm50.4$$^{a}$ & \citep{2015MNRAS.450L..16M}\\
		$2.34$    & $222.0\pm7.0$$^{c}$ &  \citep{2015AA...574A..59D}\\
		$2.36$    & $227.0\pm8.0$$^{c}$ &  \citep{2014JCAP...05..027F}\\
		\hline\hline
	\end{tabular}
	\begin{itemize}	
		\footnotesize
		\item[a] Cosmic chronometric method.
		\item[b] BAO signal in galaxy distribution.
		\item[c] BAO signal in Ly$\alpha$ forest distribution alone, or cross-correlated with quasars.
	\end{itemize}
\end{table}

Gaussian process has been extensively used for cosmological applications, such as constraint on $H_{0}$ \citep{2018JCAP...04..051G,2018ApJ...856....3Y,2019ApJ...886L..23L,2020ApJ...895L..29L} and comparison of cosmological models \citep{2018JCAP...02..034M}. Here, we only give a brief introduction to GP method. A more detailed explanation can be discovered from the literature \citep{2006gpml.book.....R,2018arXiv180702811F,SCHULZ20181}. In this work, GP regression is implemented by employing the package \emph{scikit-learn} \footnote{https://scikit-learn.org} \citep{scikit-learn} in the Python environment. It will reconstruct a continuous function $f(x)$ that is the best representative of a discrete set of measurements $f(x_{i})\pm\sigma_{i}$ at $x_{i}$, where $i$ = 1,2,..., $N$ and $\sigma_{i}$ is the 1$\sigma$ error. The GP method assumes that the value of $f(x_{i})$ at any position $x_{i}$ is random that follows a Gaussian distribution with expectation $\mu(x)$ and standard deviation $\sigma(x)$. They can be determined from observational data through a defined covariance function $k(x,x_{i})$ or kernel function
\begin{eqnarray}
	\label{eq:gp_mu}
	\mu(x) &=& \sum_{i,j = 1}^{N} k(x, x_{i})(M^{-1})_{ij}f(x_{j}),
\end{eqnarray}
and
\begin{eqnarray}
	\label{eq:gp_sigma}
	\sigma(x) &=& k(x, x_{i}) - \sum_{i,j = 1}^{N} k(x, x_{i})(M^{-1})_{ij}k(x_{j},x),
\end{eqnarray}
where the matrix $M_{ij} = k(x_{i}, x_{j}) + c_{ij}$ and $c_{ij}$ is the covariance matrix of the observed data. For the correlated measurements, the covariance matrix $c_{ij}$ is given by equation (\ref{eq:matrix}). For uncorrelated data, it can be simplified as $diag(\sigma^{2}_{i})$. Equations (\ref{eq:gp_mu}) and (\ref{eq:gp_sigma}) specify the posterior distribution of the extrapolated points.

For a given data-set ($x_{i}, y_{i}$), considering a suitable GP kernel, it is straightforward to derive the continuous function $f(x)$ which used to obtain the value of $H_{0}$, i.e. $f(0)$. In this work, we consider three kernels to illustrate the ``model dependence" of our results. The usual one is the Mat\'{e}rn kernel, whose form can be written as
\begin{eqnarray}
	\label{eq:matern}
	k(x,\tilde{x}) = \sigma^{2}_{f}(1+\frac{\sqrt{3}|x - \tilde{x}|}{l}) \textnormal{exp} (-\frac{\sqrt{3}|x - \tilde{x}|}{l}),
\end{eqnarray}
where, parameters $\sigma_{f}$ and $l$ control the strength of the correlation of the function value and the coherence length of the correlation in $x$, respectively. The other two GP kernels adopted to examine the ``model dependence" of our results are the Radial Basis Function kernel (RBF) or Gaussian kernel and the Rational Quadradtic kernel.
More detailed information about covariance functions for Gaussian process can be found in chapter 4 of the book \citep{2006gpml.book.....R}. Parameters $\sigma_f$ and $l$ are optimized for the observed data, $f(x_i)\pm\sigma_i$, by minimizing the log marginal likelihood function \citep{2012JCAP...06..036S}
\begin{eqnarray}\label{likelihood}
	\ln\mathcal{L} &=& -\frac{1}{2}\sum_{i,j=1}^N[f(x_i)-\mu(x_i)](M^{-1})_{ij}[f(x_j)-\mu(x_j)] \nonumber \\
	&-&\frac{1}{2}
	\ln|M|-\frac{1}{2}N\ln{2\pi},
\end{eqnarray}
where $|M|$ is the determinant of $M_{ij}$.
\begin{figure}
	\includegraphics[width=0.45\textwidth]{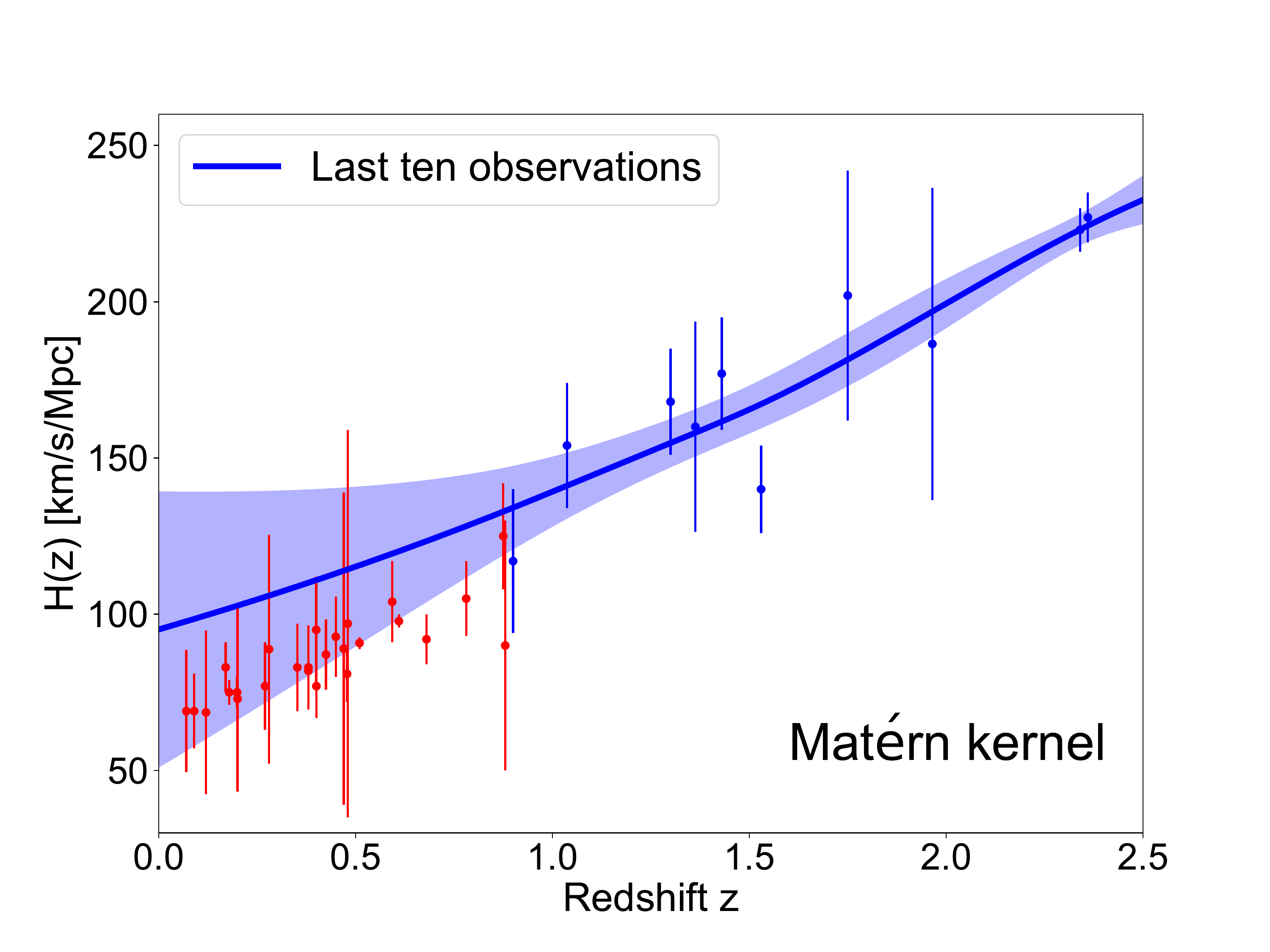}
	\caption{\label{Fig2} Reconstructed result from the last ten $H(z)$ data adopting the GP method with Mat\'{e}rn kernel. }
\end{figure}

The cumulative method used in this paper, is different from the partition method adopted in previous works \citep{2020JCAP...12..018G,2020PhRvD.102b3520K,2021ApJ...912..150D,2021MNRAS.501.3421K,2021PDU....3200814S}. The main distinction between these two methods is that the data of previous bin still used in the next bin in the cumulative method. The next bin only has one more high-redshift data than the previous bin, which means that the main difference of derived $H_{0}$ from the two bins is due to the high-redshift data. As a consequence, it makes more reasonable to write the result as $H_{0}$($z_{max}$), where $z_{max}$ represents the bin cutoff redshift. In practice, the first bin includes the first five $H(z)$ data, which has a cutoff redshift of 0.18. After that, adding one data every time until all data is included. There are 32 bins in total. Therefor we obtain 32 $H_{0}$($z_{max}$) values. A schematic diagram of the cumulative method is shown in Figure \ref{Fig1}. In addition, we also make an attempt by combining the partition method and the GP method, which is illustrated in Figure \ref{Fig2}. In this figure, the last ten data points are used to reconstruct the $H(z)$ evolution. We can see that the reconstructed $H(z)$ function generally agrees with the low-redshift observations at $1\sigma$ level, but the theoretical predictions deviate from observations and the $1\sigma$ error is too large. The main reason is that the GP method can only give a reasonable estimate within a certain range.  Therefore, a continuous $H(z)$ measurement is required. This is also the reason to adopt the cumulative binning method.

It is worth to note that $H_{0}$ represents the Hubble expansion rate at $z=0$, and $H_{0}(z_{max})$ is the value of $H_0$ derived from a data set with maximal redshift $z_{\rm max}$. Therefore, the values obtained from observational data at different redshift intervals are not a constant, implying that the expansion history of the universe may be not smooth.

\section{Results}
First, we compare the three GP kernels using the total $H(z)$ data and find that the choice of kernels has a little effect. Therefore, in below analysis 
we employ the Mat\'{e}rn kernel, which has been widely used.

\begin{figure}
	\includegraphics[width=0.45\textwidth]{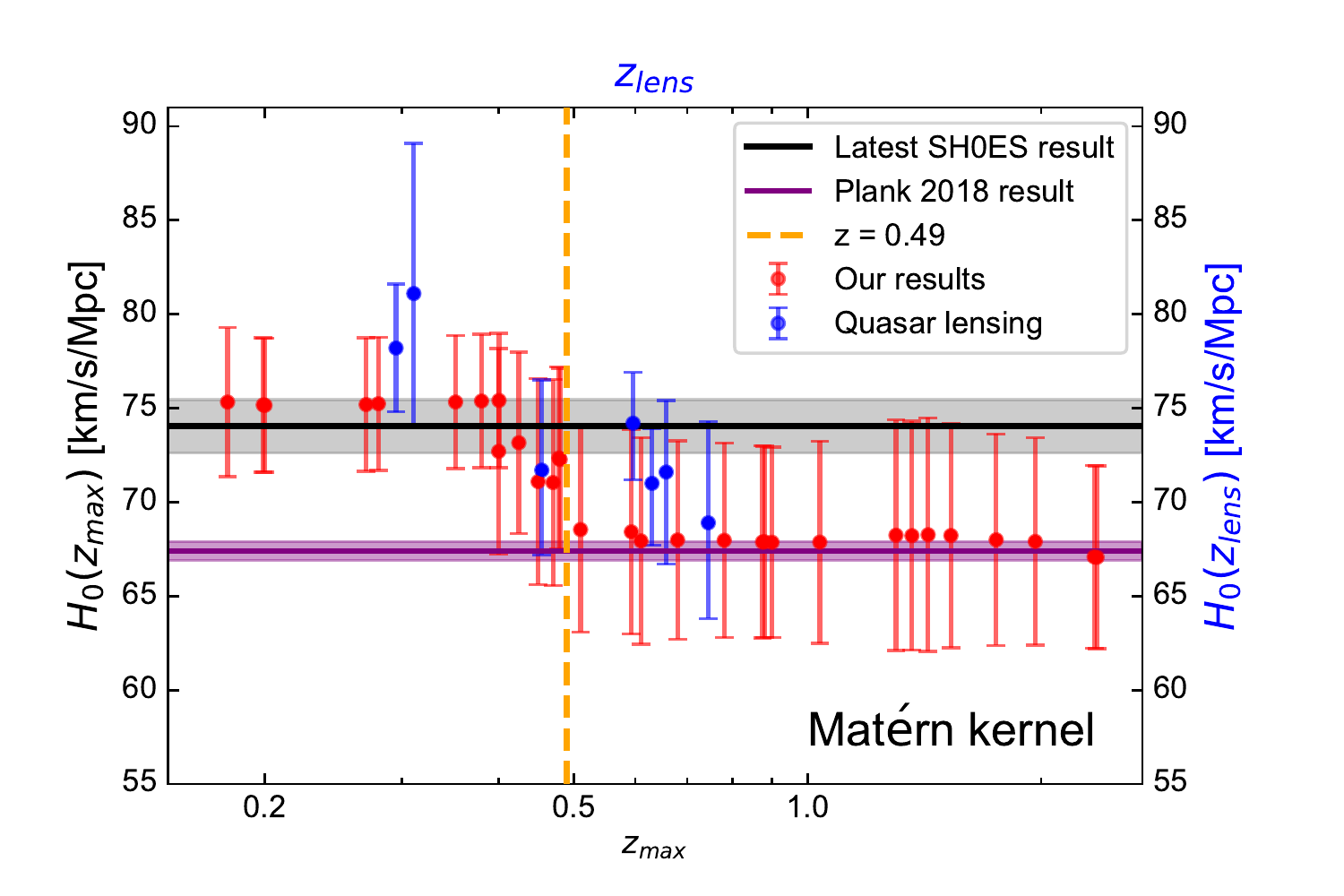}
	\caption{\label{Fig4} Predictions of $H_{0}(z_{max})$ adopting the Mat\'{e}rn kernel from the 36 $H(z)$ data binned by the cumulative method. $H_{0}(z_{max})$  is the value of $H_0$ derived from a data set with maximal redshift $z_{\rm max}$. Red points are the predictions of $H_{0}(z_{max})$ based on our analyses. The black and purple regions correspond to the results of SH0ES and \emph{Planck} collaborations. Orange dotted line ($z$ = 0.49) is the transition redshift. We also show the $H_0$ results derived from quasar lens observations with blue points in ($z_{lens}$, $H_{0}(z_{lens})$) coordinates.}
\end{figure}

\begin{figure}
	\includegraphics[width=0.45\textwidth]{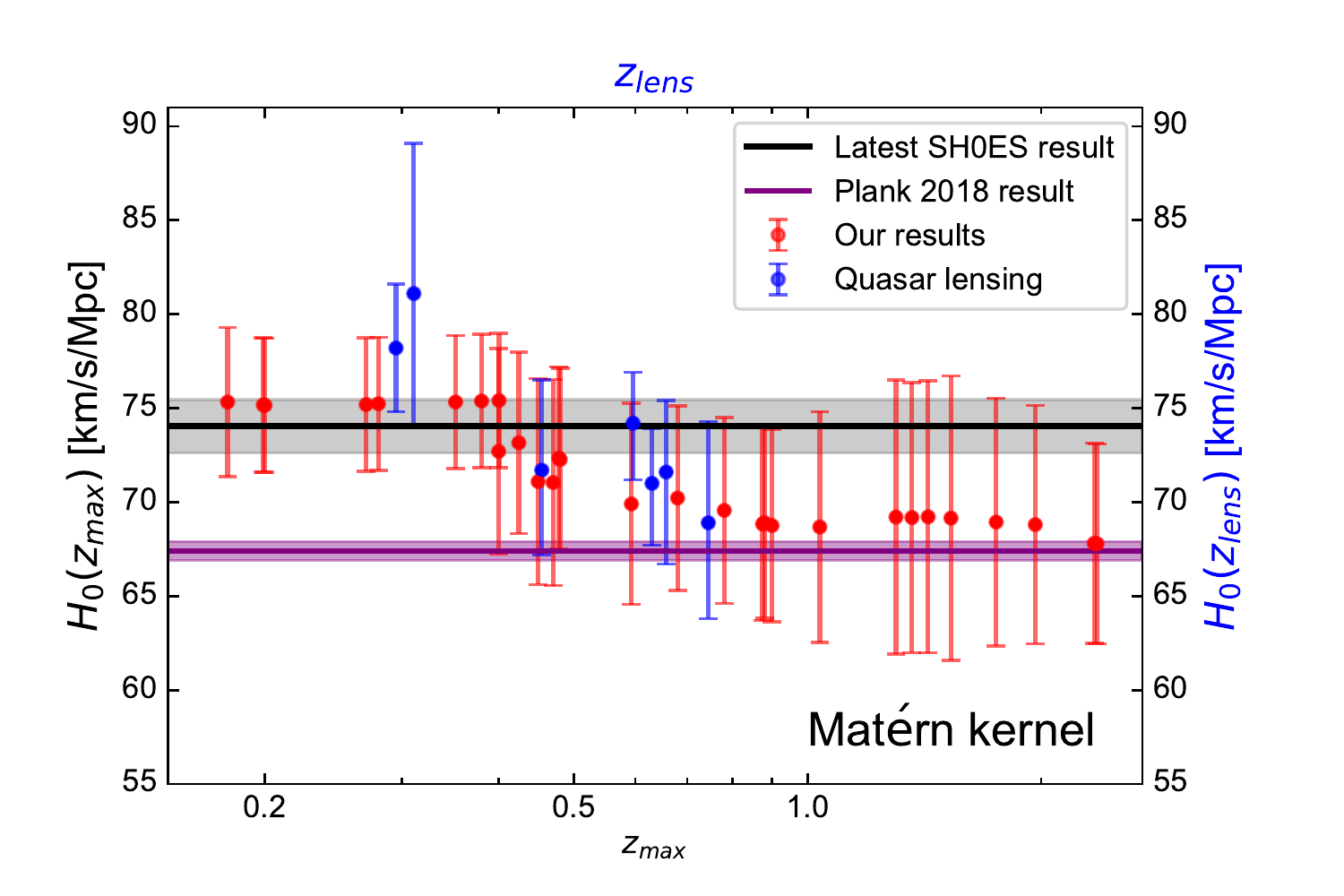}
	\caption{\label{Fig5} Predictions of $H_{0}(z_{max})$ adopting the Mat\'{e}rn kernel from 31 CCs data binned by the cumulative method. Red points are the predictions of $H_{0}(z_{max})$ based on our analyses. The black and purple regions correspond to the results of SH0ES and \emph{Planck} collaborations. We also display the $H_0$ results derived from quasar lens observations with blue points in ($z_{lens}$, $H_{0}(z_{lens})$) coordinates.}
\end{figure}

Figure \ref{Fig4} shows the value of $H_{0}(z_{max})$ from the cumulative method . We find that at early time, the values of $H_{0}$ are consistent with the \emph{Planck} result. The values at late time are consistent with the result of SH0ES collaboration. It is obvious that a $H_{0}$ transition occurred at $z\sim$0.49. The transition redshift is taken as the mean value of 0.48 and 0.51, which correspond to the maximal redshifts of the 15th bin and the 16th bin respectively. In addition, we show the values of $H_0$ provided by the H0LiCOW collaboration at different lens redshifts \citep{2020A&A...639A.101M} for comparison and find that it is consistent with our results in the 1$\sigma$ range. Recently, \citet{2021PhRvL.126w1101K} argued that $H(z)$ parameters obtained from different methods (CC, redshift drift, gravitational waves and BAO) actually describe different quantities if we are not in an FLRW universe. While they show that CC data will indeed always measure the large-scale expansion rate, this is for instance not the case for BAO. Considering this reason, we also show the results from the 31 $H(z)$ points from CC method in Figure \ref{Fig5}. We can find that the results of removing BAO data are slightly altered, and the $H_{0}$ transition still appears. From Figure \ref{Fig5}, it is not easy to point out a transition redshift of $H_{0}$, but it occurs in redshift interval 0.40 $<z<$ 0.88. This suggests that the $H_{0}$ transition is not caused by the BAO data.

It is necessary to explain why we consider $z_{lens}$ and $z_{max}$ as equivalent redshifts. With observed time delay $\Delta \tau_{obs}$ and lens mass model, $H_{0}$ can be inferred. The observed time delay is owing to the geometrical path length difference caused by the gravitational potential of the lens, which is related to the path of the light rays from the vicinity of the lens to the observer \citep{2002LNP...608.....C}. The time delay distance $D_{\Delta t}$ inferred from $\Delta \tau_{obs}$ is actually a combination of angular diameter distances: \citep{2020MNRAS.498.1420W}
\begin{eqnarray}\label{Dt}
	D_{\Delta t} \equiv (1+z_{lens})\frac{D_{d}D_{s}}{D_{ds}},
\end{eqnarray}
where $z_{lens}$ is the lens redshift, $D_{s}$ is the angular diameter distance to the source, $D_{d}$ is the angular diameter distance to the lens, and $D_{ds}$ is the angular diameter distance between the source and the lens. The time delay distance is primarily sensitive to $H_{0}$, with weak dependence on other cosmological parameters. In the flat $\Lambda$CDM model, $D_{ds}$ can be given as \citep{2018SSRv..214...91S}
\begin{eqnarray}\label{Dds}
	D_{ds} = \frac{c}{H_{0}(1+z_s)}\int_{z_{lens}}^{z_s} (\Omega_{m}(1+z')^{3}+ \Omega_{\Lambda})^{-1/2} dz',
\end{eqnarray}
here, $z_s$ is the redshift of source. According to equation (\ref{Dds}), equation (\ref{Dt}) can be rewritten as
\begin{eqnarray}\label{dd}
	D_{\Delta t} \equiv D_{d}(1+z_{lens})(1+\varepsilon),
\end{eqnarray}
where $\varepsilon$ = $\int_{0}^{z_{lens}} (\Omega_{m}(1+z')^{3}+ \Omega_{\Lambda})^{-1/2}dz'$/$\int_{z_{lens}}^{z_s} (\Omega_{m}(1+z')^{3}+ \Omega_{\Lambda})^{-1/2}dz'$ is associated with $z_s$, $z_{lens}$ and cosmological parameters. Fixed $\Omega_{m}$ = 0.3, the $\varepsilon$ values derived from six observed lens \citep{2020MNRAS.498.1420W} are in the interval (0.35, 1.23). Therefore $D_{\Delta t}$ which used to estimate $H_{0}$ is mainly contributed by the first two items of equation (\ref{dd}) and mainly related to $z_{lens}$.


Without increasing the parameter spaces and proposing a new cosmic model, our findings are simultaneously in line with the observations including the \emph{Planck}, SH0ES and quasar lens. In other words, our findings could effectively alleviate the Hubble crisis. To evaluate how much out results reduce the $H_{0}$ crisis, we compute the percentage of reduction (\%Reduc= $C_{\rm low} \times C_{\rm high}$), taking into account the local and high redshift values of $H_{0}$. The $C_{low}$ and $C_{high}$ are the degree of agreement between our results and the $H_0$ values from SH0ES and \emph{Planck} collaborations. They can be written as
\begin{eqnarray}
	C_{\rm low}(\sigma) &=& \frac{\bar H_{\rm 0,low} - H_{\rm 0, SH0ES}}{\sqrt{\bar \sigma_{\rm low}^{2}+\sigma_{\rm SH0ES}^{2}}}, \nonumber \\
	C_{\rm high}(\sigma) &=& \frac{\bar H_{\rm 0,high} - H_{\rm 0, Planck}}{\sqrt{\bar \sigma_{\rm high}^{2} + \sigma_{\rm Planck}^{2}}},
	\label{xfit}
\end{eqnarray}
where $H_{\rm 0, SH0ES}$ ($\sigma_{\rm SH0ES}$) and $H_{\rm 0, Planck}$ ($\sigma_{\rm Planck}$) are the $H_0$ values (1$\sigma$ errors) obtained by the SH0ES collaboration and the \emph{Planck} collaboration. The $\bar H_{\rm 0,low}$ ($\bar \sigma_{\rm low}$) and $\bar H_{\rm 0,high}$ ($\bar \sigma_{\rm high}$) are the mean value $H_{0}$ (1$\sigma$ errors) derived from GP method. The $\bar H_{\rm 0,low}$ is taken as the average of 6 low-redshift bins, and the $\bar H_{\rm 0,high}$ is calculated as the mean of 17 high-redshift bins. The relieved degree of $H_{0}$ crisis is estimated as 69.03\%.

In addition, we give the results in the standard $\Lambda$CDM model and $w$CDM model as a comparison. 
The best fits of $H_{0}$ are achieved by minimizing the value of
\begin{eqnarray}
	\chi^{2}_{fit} = \sum_{i=1}^{N}\frac{[H_{\rm obs}(z_{i}) - H_{\rm th}(H_{0},z_{i})]^{2}}{\sigma_{i}^{2}},
	\label{chi_mc}
\end{eqnarray}
here $H_{\rm obs}(z_{i})$ and $\sigma_{i}(z_{i})$ are the Hubble parameter and the corresponding 1$\sigma$ error. $H_{\rm th}(H_{0},z_{i})$ represents the theoretical Hubble parameter in a cosmological model, which can be given by
\begin{eqnarray}
	H_{\rm th} = H_{0}\times E(\Omega_{i}, z_{i}),
	\label{eq:Hz}
\end{eqnarray}
where $\Omega_{i}$ represents the extra cosmological parameters $\Omega_{m}$ and $w_{0}$. We marginalize them in a large range ($0<\Omega_{m}<1$, $-1.5<w_{0}<-0.5$). The likelihood analysis is performed employing a Bayesian Monte Carlo Markov Chain \citep[MCMC;][]{2013PASP..125..306F} with the \emph{emcee} (https://emcee.readthedocs.io/en/stable/) package.

\begin{figure}
	\includegraphics[width=0.45\textwidth]{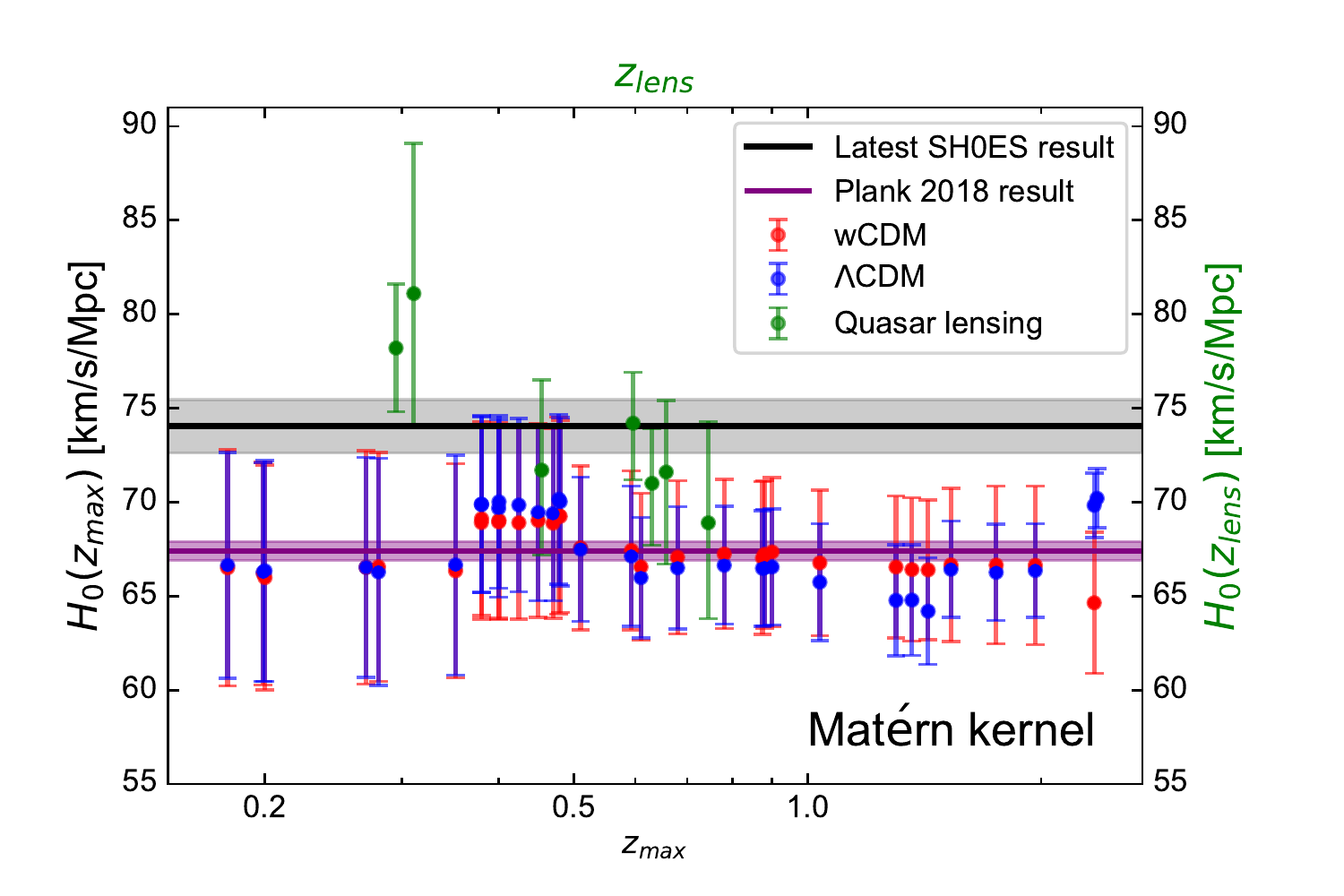}
	\caption{ \label{Fig6} The best fitting results of $H_{0}(z_{max})$ in the $\Lambda$CDM model and \emph{w}CDM model from $H(z)$ data binned by the cumulative method. The black and purple regions show the results of SH0ES and \emph{Planck} collaborations. Blue points and red points are the results derived in the $\Lambda$CDM model and \emph{w}CDM model, respectively. We also display the $H_0$ results derived from quasar lens observations with green points in ($z_{lens}$, $H_{0}(z_{lens})$) coordinates.}
\end{figure}

The results in the $\Lambda$CDM and $w$CDM models are displayed in Figure \ref{Fig6} using the same bin method above. It is worth noting that the $H_{0}$ transition disappears in these two models, but prefers a low value. But there is a bump in the redshift range (0.4, 0.5). The $H_0$ at low redshifts deviates from the measurements from time-delay cosmography of lensed quasars. So the difference is primarily caused by the first few bins.

The values of $H_{0}$ obtained by the GP method are larger but still in line with that of the $\Lambda$CDM model and the $w$CDM model within $1\sigma$ range. The main reason for this discrepancy may be caused by the relatively small data size of these bins. The number of parameters to be fitted in these two models is close to the bin size.  In addition, there should be the influence of the extra cosmological parameter $\Omega_{m}$ and $w_{0}$. Although we have integrated the extra cosmological parameters $\Omega_{m}$ and $w_{0}$ within a large range ($0<\Omega_{m}<1$, $-1.5<w_{0}<-0.5$), this process cannot completely eliminate their influence on the $H_{0}$ fitting result. Fortunately, the GP method does not need to consider this problem. To improve the accuracy of the fitting results at low redshifts, more low-redshift $H(z)$ data is needed to weaken the influence of increasing the parameter space on the fitting results.

In summary, we find a late-time transition of $H_{0}$, i.e., $H_0$ changes from a low value to a high one from early to late cosmic time, by investigating the $H(z)$ data based on a model-independent way and a cumulative binning method. Our finding effectively alleviates the Hubble crisis by 69.03\%. Our results are also consistent with the descending trend of $H_0$ measured by time-delay cosmography of lensed quasars at 1$\sigma$ confidence level.

\section{Discussion}
There has been many researches which dedicate to find out what causes the Hubble crisis, but so far no convincing explanation. Possible systematics in the Planck observations and the Hubble Space Telescope (HST) measurements have been ruled out  \citep{2017A&A...607A..95P,2018ApJ...867..108J,2019ApJ...876...85R,2019MNRAS.484L..64S,2020A&A...641A...1P,2020A&A...641A...7P,2020A&A...644A.176R,2022MNRAS.514.4620D,2022ApJ...934L...7R}. Hence, many researchers prefer to believe that the Hubble crisis may be caused by new physics beyond the $\Lambda$CDM model \citep{2020NatRP...2...10R}. Many possibilities have been proposed in the literature including but not limited to a phantom dark energy component \citep{2016PhLB..761..242D}, extra relativistic species in the early Universe \citep{2016JCAP...10..019B}, interactions between dark matter and dark energy \citep{2020PhRvD.102d3517C}, interactions between dark matter and dark radiation \citep{2017PhLB..768...12K}, vacuum energy interacting with matter and radiation \citep{2021JCAP...07..005G}, decaying dark matter \citep{2020MNRAS.497.1757H,2020JCAP...07..026P}, early recombination \citep{2021PhRvD.103h3507S}, an early dark energy component \citep{2019PhRvL.122v1301P}, modified gravity \citep{2020FoPh...50..893C,2021MNRAS.500.1795B}, etc. Until now, the mainstream is to propose an improved model based on the $\Lambda$CDM model, and then alleviate the Hubble crisis by fitting CMB or local data. After that, systematic comparison and discussion of different models which have been proposed to resolve the Hubble crisis do not give a ``successful" solution, but explicitly provide guidance to model builders \citep{2019JCAP...02..054G,2020PhRvD.101d3533K,2022PhR...984....1S}. \citet{2022PhRvD.105b1301C} proposed to use a global parameterization based on the cosmic age to consistently use the cosmic chronometers data beyond the Taylor expansion domain and without the input of a sound-horizon prior. Both the early-time and late-time scenarios are therefore largely ruled out. From the recent work, it is difficult to construct an early-time resolution to the Hubble crisis \citep{2021PhRvD.104f3524V}.

Unlike previous works \citep{2019JCAP...12..035K,2019MNRAS.483.4803L,2020A&A...643A.165B}, we want to use model-independent methods to find some anomalous behaviors from the Hubble parameter data that could be used to explain the Hubble crisis. Meanwhile, some recent work suggests that the $H_{0}$ descending trend with redshift can be used to explain the Hubble crisis. So we focus on the local data. We find a late-time transition of $H_{0}$ from the $H(z)$ data, using the GP method and the cumulative binning method. As the cutoff redshift ($z_{max}$) of the dataset decreases, the $H_{0}$ value transforms from a low value to a high value. Without proposing a new cosmological model, this finding can be used to relax the Hubble crisis with a mitigation level of around 70\%. If the late-time transition behavior of $H_{0}$ is true, it suggests that the Hubble crisis is most likely due to the new physics beyond the standard $\Lambda$CDM model.

What causes the transition? The local void \citep{2008GReGr..40..451E,2008JCAP...04..003G,2013ApJ...775...62K,Wang2013} and modified gravity \citep{2020FoPh...50..893C,2021MNRAS.500.1795B} might be possible solutions. The local void has low matter density, which in turn increases the value of $H_0$. Local void as a late-time solution \citep{2008JCAP...04..003G,2013ApJ...775...62K} has been disfavored by the SNe Ia data \citep{2019ApJ...875..145K,2020MNRAS.491.2075L,2021PhRvD.103l3539C}, but can not completely be ruled out. In addition, there is some evidences supporting the existence of local void model \citep{2022arXiv220113384K}. Adopting the spatial distribution of galaxy clusters to map the matter density distribution in the local Universe, \citet{2020A&A...633A..19B} find a local underdensity in the cluster distribution. Inside this underdensity, the observed Hubble parameter will be larger by about 5.50\%, which can be used to explain the Hubble crisis. Except the local void models, some modified cosmological models could also be used to explain our finding, i.e., a quintessence field which trasitions from a matter-like to a cosmological constant behavior between recombination and the present time \citep{2019PDU....2600385D}.

At present, the error of the derived $H_{0}$ is large, which is mainly limited by the sample size and observational error. Due to the lack of $H(z)$ data at high redshifts, whether the high-redshift result is still consistent with the \emph{Planck} result is unclear. Validation of our findings requires more high-quality $H(z)$ observations, especially high-redshift data. At the same time, we also expect that more $H_{0}$ can be derived from the quasar lens, which can also be used to verify that the late-time transition of $H_{0}$. From this point of view, future missions like the large Synoptic Survey Telescope \citep{2009arXiv0912.0201L}, Euclid \citep{2011arXiv1110.3193L} and Wide Field Infra Red Survey Telescope \citep{2013arXiv1305.5422S} will achieve more high-quality $H(z)$ data which can improve our research. In addition, measuring $H_{0}$ from gravitational wave standard sirens and fast radio bursts could also help us checking out our findings \citep{2017Natur.551...85A,2018Natur.562..545C,2019NatAs...3..384C,2019PhRvL.122f1105F,2022MNRAS.511..662H,2022MNRAS.515L...1W}.

In summary, our findings suggest that the use of late-time solutions, like local void models, to resolve the Hubble crisis is worth to investigate. In future, we plan to explore the model-independent trend of $H_{0}$ using other observational data, such as supernovae, gamma-ray bursts, and quasars, adopting the cumulative method. Similar $H_0$ evolution is found from model-independent analyses of the SNe Ia Pantheon sample \citep{Hu&Wangprep}.

\section*{Acknowledgements}
We thank the anonymous referee for constructive comments. We thank Zuo-Lin Tu, Sofie Marie Koksbang, Sunny Vagnozzi, Eoin $\rm \acute{O}$ Colg$\rm \acute{a}$in for helpful discussion and comments. This work was supported by the National Natural Science
Foundation of China (grant No. U1831207), the China Manned Spaced Project (CMS-CSST-2021-A12) and Jiangsu Funding Program for Excellent Postdoctoral Talent (20220ZB59).

\section*{DATA AVAILABILITY}
The data used in the paper are publicly available in Table 1.

\bibliographystyle{mnras}
\bibliography{mnras_mg} 

\begin{thebibliography}{}
\makeatletter
\relax
\def\mn@urlcharsother{\let\do\@makeother \do\$\do\&\do\#\do\^\do\_\do\%\do\~}
\def\mn@doi{\begingroup\mn@urlcharsother \@ifnextchar [ {\mn@doi@}
  {\mn@doi@[]}}
\def\mn@doi@[#1]#2{\def\@tempa{#1}\ifx\@tempa\@empty \href
  {http://dx.doi.org/#2} {doi:#2}\else \href {http://dx.doi.org/#2} {#1}\fi
  \endgroup}
\def\mn@eprint#1#2{\mn@eprint@#1:#2::\@nil}
\def\mn@eprint@arXiv#1{\href {http://arxiv.org/abs/#1} {{\tt arXiv:#1}}}
\def\mn@eprint@dblp#1{\href {http://dblp.uni-trier.de/rec/bibtex/#1.xml}
  {dblp:#1}}
\def\mn@eprint@#1:#2:#3:#4\@nil{\def\@tempa {#1}\def\@tempb {#2}\def\@tempc
  {#3}\ifx \@tempc \@empty \let \@tempc \@tempb \let \@tempb \@tempa \fi \ifx
  \@tempb \@empty \def\@tempb {arXiv}\fi \@ifundefined
  {mn@eprint@\@tempb}{\@tempb:\@tempc}{\expandafter \expandafter \csname
  mn@eprint@\@tempb\endcsname \expandafter{\@tempc}}}

\bibitem[\protect\citeauthoryear{{Abbott} et~al.,}{{Abbott}
  et~al.}{2017}]{2017Natur.551...85A}
{Abbott} B.~P.,  et~al., 2017, \mn@doi [\nat] {10.1038/nature24471}, \href
  {https://ui.adsabs.harvard.edu/abs/2017Natur.551...85A} {551, 85}

\bibitem[\protect\citeauthoryear{{Alam}, {Ata}, {Bailey}, {Beutler}, {Bizyaev}
  \& {et al.}}{{Alam} et~al.}{2017}]{2017MNRAS.470.2617A}
{Alam} S.,  {Ata} M.,  {Bailey} S.,  {Beutler} F.,  {Bizyaev} D.,   {et al.}
  2017, \mn@doi [\mnras] {10.1093/mnras/stx721}, \href
  {https://ui.adsabs.harvard.edu/abs/2017MNRAS.470.2617A} {470, 2617}

\bibitem[\protect\citeauthoryear{{Amati}, {D'Agostino}, {Luongo}, {Muccino}  \&
  {Tantalo}}{{Amati} et~al.}{2019}]{2019MNRAS.486L..46A}
{Amati} L.,  {D'Agostino} R.,  {Luongo} O.,  {Muccino} M.,   {Tantalo} M.,
  2019, \mn@doi [\mnras] {10.1093/mnrasl/slz056}, \href
  {https://ui.adsabs.harvard.edu/abs/2019MNRAS.486L..46A} {486, L46}

\bibitem[\protect\citeauthoryear{{Benetti}, {Capozziello}  \&
  {Lambiase}}{{Benetti} et~al.}{2021}]{2021MNRAS.500.1795B}
{Benetti} M.,  {Capozziello} S.,   {Lambiase} G.,  2021, \mn@doi [\mnras]
  {10.1093/mnras/staa3368}, \href
  {https://ui.adsabs.harvard.edu/abs/2021MNRAS.500.1795B} {500, 1795}

\bibitem[\protect\citeauthoryear{{Benisty} \& {Staicova}}{{Benisty} \&
  {Staicova}}{2021}]{2021A&A...647A..38B}
{Benisty} D.,  {Staicova} D.,  2021, \mn@doi [\aap]
  {10.1051/0004-6361/202039502}, \href
  {https://ui.adsabs.harvard.edu/abs/2021A&A...647A..38B} {647, A38}

\bibitem[\protect\citeauthoryear{{Bernal}, {Verde}  \& {Riess}}{{Bernal}
  et~al.}{2016}]{2016JCAP...10..019B}
{Bernal} J.~L.,  {Verde} L.,   {Riess} A.~G.,  2016, \mn@doi [\jcap]
  {10.1088/1475-7516/2016/10/019}, \href
  {https://ui.adsabs.harvard.edu/abs/2016JCAP...10..019B} {2016, 019}

\bibitem[\protect\citeauthoryear{{Bilicki} \& {Seikel}}{{Bilicki} \&
  {Seikel}}{2012}]{2012MNRAS.425.1664B}
{Bilicki} M.,  {Seikel} M.,  2012, \mn@doi [\mnras]
  {10.1111/j.1365-2966.2012.21575.x}, \href
  {https://ui.adsabs.harvard.edu/abs/2012MNRAS.425.1664B} {425, 1664}

\bibitem[\protect\citeauthoryear{{Birrer} et~al.,}{{Birrer}
  et~al.}{2020}]{2020A&A...643A.165B}
{Birrer} S.,  et~al., 2020, \mn@doi [\aap] {10.1051/0004-6361/202038861}, \href
  {https://ui.adsabs.harvard.edu/abs/2020A&A...643A.165B} {643, A165}

\bibitem[\protect\citeauthoryear{{B{\"o}hringer}, {Chon}  \&
  {Collins}}{{B{\"o}hringer} et~al.}{2020}]{2020A&A...633A..19B}
{B{\"o}hringer} H.,  {Chon} G.,   {Collins} C.~A.,  2020, \mn@doi [\aap]
  {10.1051/0004-6361/201936400}, \href
  {https://ui.adsabs.harvard.edu/abs/2020A&A...633A..19B} {633, A19}

\bibitem[\protect\citeauthoryear{{Cai}, {Ding}, {Guo}, {Wang}  \& {Yu}}{{Cai}
  et~al.}{2021}]{2021PhRvD.103l3539C}
{Cai} R.-G.,  {Ding} J.-F.,  {Guo} Z.-K.,  {Wang} S.-J.,   {Yu} W.-W.,  2021,
  \mn@doi [\prd] {10.1103/PhysRevD.103.123539}, \href
  {https://ui.adsabs.harvard.edu/abs/2021PhRvD.103l3539C} {103, 123539}

\bibitem[\protect\citeauthoryear{{Cai}, {Guo}, {Wang}, {Yu}  \& {Zhou}}{{Cai}
  et~al.}{2022}]{2022PhRvD.105b1301C}
{Cai} R.-G.,  {Guo} Z.-K.,  {Wang} S.-J.,  {Yu} W.-W.,   {Zhou} Y.,  2022,
  \mn@doi [\prd] {10.1103/PhysRevD.105.L021301}, \href
  {https://ui.adsabs.harvard.edu/abs/2022PhRvD.105b1301C} {105, L021301}

\bibitem[\protect\citeauthoryear{{Capozziello}, {Benetti}  \&
  {Spallicci}}{{Capozziello} et~al.}{2020}]{2020FoPh...50..893C}
{Capozziello} S.,  {Benetti} M.,   {Spallicci} A. D.~A.~M.,  2020, \mn@doi
  [Foundations of Physics] {10.1007/s10701-020-00356-2}, \href
  {https://ui.adsabs.harvard.edu/abs/2020FoPh...50..893C} {50, 893}

\bibitem[\protect\citeauthoryear{{Cawthon} et~al.,}{{Cawthon}
  et~al.}{2022}]{2022MNRAS.513.5517C}
{Cawthon} R.,  et~al., 2022, \mn@doi [\mnras] {10.1093/mnras/stac1160}, \href
  {https://ui.adsabs.harvard.edu/abs/2022MNRAS.513.5517C} {513, 5517}

\bibitem[\protect\citeauthoryear{{Chen}}{{Chen}}{2019}]{2019NatAs...3..384C}
{Chen} H.-Y.,  2019, \mn@doi [Nature Astronomy] {10.1038/s41550-019-0776-1},
  \href {https://ui.adsabs.harvard.edu/abs/2019NatAs...3..384C} {3, 384}

\bibitem[\protect\citeauthoryear{{Chen}, {Fishbach}  \& {Holz}}{{Chen}
  et~al.}{2018}]{2018Natur.562..545C}
{Chen} H.-Y.,  {Fishbach} M.,   {Holz} D.~E.,  2018, \mn@doi [\nat]
  {10.1038/s41586-018-0606-0}, \href
  {https://ui.adsabs.harvard.edu/abs/2018Natur.562..545C} {562, 545}

\bibitem[\protect\citeauthoryear{{Cheng}, {Ma}, {Wu}, {Zhang}  \&
  {Chen}}{{Cheng} et~al.}{2020}]{2020PhRvD.102d3517C}
{Cheng} G.,  {Ma} Y.-Z.,  {Wu} F.,  {Zhang} J.,   {Chen} X.,  2020, \mn@doi
  [\prd] {10.1103/PhysRevD.102.043517}, \href
  {https://ui.adsabs.harvard.edu/abs/2020PhRvD.102d3517C} {102, 043517}

\bibitem[\protect\citeauthoryear{{Colg{\'a}in}, {Sheikh-Jabbari}, {Solomon},
  {Bargiacchi}, {Capozziello}, {Dainotti}  \& {Stojkovic}}{{Colg{\'a}in}
  et~al.}{2022}]{2022arXiv220310558C}
{Colg{\'a}in} E.~{\'O}.,  {Sheikh-Jabbari} M.~M.,  {Solomon} R.,  {Bargiacchi}
  G.,  {Capozziello} S.,  {Dainotti} M.~G.,   {Stojkovic} D.,  2022, arXiv
  e-prints, \href {https://ui.adsabs.harvard.edu/abs/2022arXiv220310558C} {p.
  arXiv:2203.10558}

\bibitem[\protect\citeauthoryear{{Courbin} \& {Minniti}}{{Courbin} \&
  {Minniti}}{2002}]{2002LNP...608.....C}
{Courbin} F.,  {Minniti} D.,  2002, {Gravitational Lensing: An Astrophysical
  Tool}.
 Vol. 608

\bibitem[\protect\citeauthoryear{{Dainotti}, {De Simone}, {Schiavone},
  {Montani}, {Rinaldi}  \& {et al.}}{{Dainotti}
  et~al.}{2021}]{2021ApJ...912..150D}
{Dainotti} M.~G.,  {De Simone} B.,  {Schiavone} T.,  {Montani} G.,  {Rinaldi}
  E.,   {et al.} 2021, \mn@doi [\apj] {10.3847/1538-4357/abeb73}, \href
  {https://ui.adsabs.harvard.edu/abs/2021ApJ...912..150D} {912, 150}

\bibitem[\protect\citeauthoryear{{Delubac}, {Bautista}, {Busca}, {Rich},
  {Kirkby}  \& {et al.}}{{Delubac} et~al.}{2015}]{2015AA...574A..59D}
{Delubac} T.,  {Bautista} J.~E.,  {Busca} N.~G.,  {Rich} J.,  {Kirkby} D.,
  {et al.} 2015, \mn@doi [\aap] {10.1051/0004-6361/201423969}, \href
  {https://ui.adsabs.harvard.edu/abs/2015A&A...574A..59D} {574, A59}

\bibitem[\protect\citeauthoryear{{Di Valentino}, {Melchiorri}  \& {Silk}}{{Di
  Valentino} et~al.}{2016}]{2016PhLB..761..242D}
{Di Valentino} E.,  {Melchiorri} A.,   {Silk} J.,  2016, \mn@doi [Physics
  Letters B] {10.1016/j.physletb.2016.08.043}, \href
  {https://ui.adsabs.harvard.edu/abs/2016PhLB..761..242D} {761, 242}

\bibitem[\protect\citeauthoryear{{Di Valentino}, {Ferreira}, {Visinelli}  \&
  {Danielsson}}{{Di Valentino} et~al.}{2019}]{2019PDU....2600385D}
{Di Valentino} E.,  {Ferreira} R.~Z.,  {Visinelli} L.,   {Danielsson} U.,
  2019, \mn@doi [Physics of the Dark Universe] {10.1016/j.dark.2019.100385},
  \href {https://ui.adsabs.harvard.edu/abs/2019PDU....2600385D} {26, 100385}

\bibitem[\protect\citeauthoryear{{Di Valentino}, {Mena}, {Pan}, {Visinelli},
  {Yang}  \& {et al.}}{{Di Valentino} et~al.}{2021}]{2021CQGra..38o3001D}
{Di Valentino} E.,  {Mena} O.,  {Pan} S.,  {Visinelli} L.,  {Yang} W.,   {et
  al.} 2021, \mn@doi [Classical and Quantum Gravity]
  {10.1088/1361-6382/ac086d}, \href
  {https://ui.adsabs.harvard.edu/abs/2021CQGra..38o3001D} {38, 153001}

\bibitem[\protect\citeauthoryear{{Enqvist}}{{Enqvist}}{2008}]{2008GReGr..40..451E}
{Enqvist} K.,  2008, \mn@doi [General Relativity and Gravitation]
  {10.1007/s10714-007-0553-9}, \href
  {https://ui.adsabs.harvard.edu/abs/2008GReGr..40..451E} {40, 451}

\bibitem[\protect\citeauthoryear{{Feeney}, {Peiris}, {Williamson}, {Nissanke},
  {Mortlock}, {Alsing}  \& {Scolnic}}{{Feeney}
  et~al.}{2019}]{2019PhRvL.122f1105F}
{Feeney} S.~M.,  {Peiris} H.~V.,  {Williamson} A.~R.,  {Nissanke} S.~M.,
  {Mortlock} D.~J.,  {Alsing} J.,   {Scolnic} D.,  2019, \mn@doi [\prl]
  {10.1103/PhysRevLett.122.061105}, \href
  {https://ui.adsabs.harvard.edu/abs/2019PhRvL.122f1105F} {122, 061105}

\bibitem[\protect\citeauthoryear{{Font-Ribera}, {Kirkby}, {Busca},
  {Miralda-Escud{\'e}}, {Ross}  \& {et al.}}{{Font-Ribera}
  et~al.}{2014}]{2014JCAP...05..027F}
{Font-Ribera} A.,  {Kirkby} D.,  {Busca} N.,  {Miralda-Escud{\'e}} J.,  {Ross}
  N.~P.,   {et al.} 2014, \mn@doi [\jcap] {10.1088/1475-7516/2014/05/027},
  \href {https://ui.adsabs.harvard.edu/abs/2014JCAP...05..027F} {2014, 027}

\bibitem[\protect\citeauthoryear{{Foreman-Mackey}, {Hogg}, {Lang}  \&
  {Goodman}}{{Foreman-Mackey} et~al.}{2013}]{2013PASP..125..306F}
{Foreman-Mackey} D.,  {Hogg} D.~W.,  {Lang} D.,   {Goodman} J.,  2013, \mn@doi
  [\pasp] {10.1086/670067}, \href
  {https://ui.adsabs.harvard.edu/abs/2013PASP..125..306F} {125, 306}

\bibitem[\protect\citeauthoryear{{Frazier}}{{Frazier}}{2018}]{2018arXiv180702811F}
{Frazier} P.~I.,  2018, arXiv e-prints, \href
  {https://ui.adsabs.harvard.edu/abs/2018arXiv180702811F} {p. arXiv:1807.02811}

\bibitem[\protect\citeauthoryear{{Freedman}, {Madore}, {Hatt}, {Hoyt}, {Jang}
  \& {et al.}}{{Freedman} et~al.}{2019}]{2019ApJ...882...34F}
{Freedman} W.~L.,  {Madore} B.~F.,  {Hatt} D.,  {Hoyt} T.~J.,  {Jang} I.~S.,
  {et al.} 2019, \mn@doi [\apj] {10.3847/1538-4357/ab2f73}, \href
  {https://ui.adsabs.harvard.edu/abs/2019ApJ...882...34F} {882, 34}

\bibitem[\protect\citeauthoryear{{Gao}, {Zhao}, {Xue}  \& {Zhang}}{{Gao}
  et~al.}{2021}]{2021JCAP...07..005G}
{Gao} L.-Y.,  {Zhao} Z.-W.,  {Xue} S.-S.,   {Zhang} X.,  2021, \mn@doi [\jcap]
  {10.1088/1475-7516/2021/07/005}, \href
  {https://ui.adsabs.harvard.edu/abs/2021JCAP...07..005G} {2021, 005}

\bibitem[\protect\citeauthoryear{{Garcia-Bellido} \&
  {Haugb{\o}lle}}{{Garcia-Bellido} \&
  {Haugb{\o}lle}}{2008}]{2008JCAP...04..003G}
{Garcia-Bellido} J.,  {Haugb{\o}lle} T.,  2008, \mn@doi [\jcap]
  {10.1088/1475-7516/2008/04/003}, \href
  {https://ui.adsabs.harvard.edu/abs/2008JCAP...04..003G} {2008, 003}

\bibitem[\protect\citeauthoryear{{Garcia-Quintero}, {Ishak}  \&
  {Ning}}{{Garcia-Quintero} et~al.}{2020}]{2020JCAP...12..018G}
{Garcia-Quintero} C.,  {Ishak} M.,   {Ning} O.,  2020, \mn@doi [\jcap]
  {10.1088/1475-7516/2020/12/018}, \href
  {https://ui.adsabs.harvard.edu/abs/2020JCAP...12..018G} {2020, 018}

\bibitem[\protect\citeauthoryear{{G{\'o}mez-Valent} \&
  {Amendola}}{{G{\'o}mez-Valent} \& {Amendola}}{2018}]{2018JCAP...04..051G}
{G{\'o}mez-Valent} A.,  {Amendola} L.,  2018, \mn@doi [\jcap]
  {10.1088/1475-7516/2018/04/051}, \href
  {https://ui.adsabs.harvard.edu/abs/2018JCAP...04..051G} {2018, 051}

\bibitem[\protect\citeauthoryear{{Guo}, {Zhang}  \& {Zhang}}{{Guo}
  et~al.}{2019}]{2019JCAP...02..054G}
{Guo} R.-Y.,  {Zhang} J.-F.,   {Zhang} X.,  2019, \mn@doi [\jcap]
  {10.1088/1475-7516/2019/02/054}, \href
  {https://ui.adsabs.harvard.edu/abs/2019JCAP...02..054G} {2019, 054}

\bibitem[\protect\citeauthoryear{{Hagstotz}, {Reischke}  \& {Lilow}}{{Hagstotz}
  et~al.}{2022}]{2022MNRAS.511..662H}
{Hagstotz} S.,  {Reischke} R.,   {Lilow} R.,  2022, \mn@doi [\mnras]
  {10.1093/mnras/stac077}, \href
  {https://ui.adsabs.harvard.edu/abs/2022MNRAS.511..662H} {511, 662}

\bibitem[\protect\citeauthoryear{{Haridasu} \& {Viel}}{{Haridasu} \&
  {Viel}}{2020}]{2020MNRAS.497.1757H}
{Haridasu} B.~S.,  {Viel} M.,  2020, \mn@doi [\mnras] {10.1093/mnras/staa1991},
  \href {https://ui.adsabs.harvard.edu/abs/2020MNRAS.497.1757H} {497, 1757}

\bibitem[\protect\citeauthoryear{{Holsclaw}, {Alam}, {Sans{\'o}}, {Lee},
  {Heitmann}  \& {et al.}}{{Holsclaw} et~al.}{2010}]{2010PhRvL.105x1302H}
{Holsclaw} T.,  {Alam} U.,  {Sans{\'o}} B.,  {Lee} H.,  {Heitmann} K.,   {et
  al.} 2010, \mn@doi [\prl] {10.1103/PhysRevLett.105.241302}, \href
  {https://ui.adsabs.harvard.edu/abs/2010PhRvL.105x1302H} {105, 241302}

\bibitem[\protect\citeauthoryear{{Horstmann}, {Pietschke}  \&
  {Schwarz}}{{Horstmann} et~al.}{2021}]{2021arXiv211103055H}
{Horstmann} N.,  {Pietschke} Y.,   {Schwarz} D.~J.,  2021, arXiv e-prints,
  \href {https://ui.adsabs.harvard.edu/abs/2021arXiv211103055H} {p.
  arXiv:2111.03055}

\bibitem[\protect\citeauthoryear{{Hu} \& {Wang}}{{Hu} \&
  {Wang}}{2022}]{Hu&Wangprep}
{Hu} J.~P.,  {Wang} F.~Y.,  2022, Manuscript in preparation

\bibitem[\protect\citeauthoryear{{Hu}, {Wang}  \& {Dai}}{{Hu}
  et~al.}{2021}]{2021MNRAS.507..730H}
{Hu} J.~P.,  {Wang} F.~Y.,   {Dai} Z.~G.,  2021, \mn@doi [\mnras]
  {10.1093/mnras/stab2180}, \href
  {https://ui.adsabs.harvard.edu/abs/2021MNRAS.507..730H} {507, 730}

\bibitem[\protect\citeauthoryear{{Inserra} et~al.,}{{Inserra}
  et~al.}{2021}]{2021MNRAS.504.2535I}
{Inserra} C.,  et~al., 2021, \mn@doi [\mnras] {10.1093/mnras/stab978}, \href
  {https://ui.adsabs.harvard.edu/abs/2021MNRAS.504.2535I} {504, 2535}

\bibitem[\protect\citeauthoryear{{Jimenez} \& {Loeb}}{{Jimenez} \&
  {Loeb}}{2002}]{2002ApJ...573...37J}
{Jimenez} R.,  {Loeb} A.,  2002, \mn@doi [\apj] {10.1086/340549}, \href
  {https://ui.adsabs.harvard.edu/abs/2002ApJ...573...37J} {573, 37}

\bibitem[\protect\citeauthoryear{{Jones} et~al.,}{{Jones}
  et~al.}{2018}]{2018ApJ...867..108J}
{Jones} D.~O.,  et~al., 2018, \mn@doi [\apj] {10.3847/1538-4357/aae2b9}, \href
  {https://ui.adsabs.harvard.edu/abs/2018ApJ...867..108J} {867, 108}

\bibitem[\protect\citeauthoryear{{Kazantzidis} \&
  {Perivolaropoulos}}{{Kazantzidis} \&
  {Perivolaropoulos}}{2020}]{2020PhRvD.102b3520K}
{Kazantzidis} L.,  {Perivolaropoulos} L.,  2020, \mn@doi [\prd]
  {10.1103/PhysRevD.102.023520}, \href
  {https://ui.adsabs.harvard.edu/abs/2020PhRvD.102b3520K} {102, 023520}

\bibitem[\protect\citeauthoryear{{Kazantzidis}, {Koo}, {Nesseris},
  {Perivolaropoulos}  \& {Shafieloo}}{{Kazantzidis}
  et~al.}{2021}]{2021MNRAS.501.3421K}
{Kazantzidis} L.,  {Koo} H.,  {Nesseris} S.,  {Perivolaropoulos} L.,
  {Shafieloo} A.,  2021, \mn@doi [\mnras] {10.1093/mnras/staa3866}, \href
  {https://ui.adsabs.harvard.edu/abs/2021MNRAS.501.3421K} {501, 3421}

\bibitem[\protect\citeauthoryear{{Keeley}, {Joudaki}, {Kaplinghat}  \&
  {Kirkby}}{{Keeley} et~al.}{2019}]{2019JCAP...12..035K}
{Keeley} R.~E.,  {Joudaki} S.,  {Kaplinghat} M.,   {Kirkby} D.,  2019, \mn@doi
  [\jcap] {10.1088/1475-7516/2019/12/035}, \href
  {https://ui.adsabs.harvard.edu/abs/2019JCAP...12..035K} {2019, 035}

\bibitem[\protect\citeauthoryear{{Keenan}, {Barger}  \& {Cowie}}{{Keenan}
  et~al.}{2013}]{2013ApJ...775...62K}
{Keenan} R.~C.,  {Barger} A.~J.,   {Cowie} L.~L.,  2013, \mn@doi [\apj]
  {10.1088/0004-637X/775/1/62}, \href
  {https://ui.adsabs.harvard.edu/abs/2013ApJ...775...62K} {775, 62}

\bibitem[\protect\citeauthoryear{{Kenworthy}, {Scolnic}  \&
  {Riess}}{{Kenworthy} et~al.}{2019}]{2019ApJ...875..145K}
{Kenworthy} W.~D.,  {Scolnic} D.,   {Riess} A.,  2019, \mn@doi [\apj]
  {10.3847/1538-4357/ab0ebf}, \href
  {https://ui.adsabs.harvard.edu/abs/2019ApJ...875..145K} {875, 145}

\bibitem[\protect\citeauthoryear{{Knox} \& {Millea}}{{Knox} \&
  {Millea}}{2020}]{2020PhRvD.101d3533K}
{Knox} L.,  {Millea} M.,  2020, \mn@doi [\prd] {10.1103/PhysRevD.101.043533},
  \href {https://ui.adsabs.harvard.edu/abs/2020PhRvD.101d3533K} {101, 043533}

\bibitem[\protect\citeauthoryear{{Ko} \& {Tang}}{{Ko} \&
  {Tang}}{2017}]{2017PhLB..768...12K}
{Ko} P.,  {Tang} Y.,  2017, \mn@doi [Physics Letters B]
  {10.1016/j.physletb.2017.02.033}, \href
  {https://ui.adsabs.harvard.edu/abs/2017PhLB..768...12K} {768, 12}

\bibitem[\protect\citeauthoryear{{Koksbang}}{{Koksbang}}{2021}]{2021PhRvL.126w1101K}
{Koksbang} S.~M.,  2021, \mn@doi [\prl] {10.1103/PhysRevLett.126.231101}, \href
  {https://ui.adsabs.harvard.edu/abs/2021PhRvL.126w1101K} {126, 231101}

\bibitem[\protect\citeauthoryear{{Krishnan} \& {Mondol}}{{Krishnan} \&
  {Mondol}}{2022}]{2022arXiv220113384K}
{Krishnan} C.,  {Mondol} R.,  2022, arXiv e-prints, \href
  {https://ui.adsabs.harvard.edu/abs/2022arXiv220113384K} {p. arXiv:2201.13384}

\bibitem[\protect\citeauthoryear{{Krishnan}, {Colg{\'a}in}, {Ruchika},
  {Sheikh-Jabbari}  \& {Yang}}{{Krishnan} et~al.}{2020}]{2020PhRvD.102j3525K}
{Krishnan} C.,  {Colg{\'a}in} E.~{\'O}.,  {Ruchika} Sen A.~A.,
  {Sheikh-Jabbari} M.~M.,   {Yang} T.,  2020, \mn@doi [\prd]
  {10.1103/PhysRevD.102.103525}, \href
  {https://ui.adsabs.harvard.edu/abs/2020PhRvD.102j3525K} {102, 103525}

\bibitem[\protect\citeauthoryear{{LSST Science Collaboration}, {Abell},
  {Allison}, {Anderson}, {Andrew}  \& {et al.}}{{LSST Science Collaboration}
  et~al.}{2009}]{2009arXiv0912.0201L}
{LSST Science Collaboration} {Abell} P.~A.,  {Allison} J.,  {Anderson} S.~F.,
  {Andrew} J.~R.,   {et al.} 2009, arXiv e-prints, \href
  {https://ui.adsabs.harvard.edu/abs/2009arXiv0912.0201L} {p. arXiv:0912.0201}

\bibitem[\protect\citeauthoryear{{Laureijs}, {Amiaux}, {Arduini},
  {Augu{\`e}res}, {Brinchmann}  \& {et al.}}{{Laureijs}
  et~al.}{2011}]{2011arXiv1110.3193L}
{Laureijs} R.,  {Amiaux} J.,  {Arduini} S.,  {Augu{\`e}res} J.~L.,
  {Brinchmann} J.,   {et al.} 2011, arXiv e-prints, \href
  {https://ui.adsabs.harvard.edu/abs/2011arXiv1110.3193L} {p. arXiv:1110.3193}

\bibitem[\protect\citeauthoryear{{Lemos}, {Lee}, {Efstathiou}  \&
  {Gratton}}{{Lemos} et~al.}{2019}]{2019MNRAS.483.4803L}
{Lemos} P.,  {Lee} E.,  {Efstathiou} G.,   {Gratton} S.,  2019, \mn@doi
  [\mnras] {10.1093/mnras/sty3082}, \href
  {https://ui.adsabs.harvard.edu/abs/2019MNRAS.483.4803L} {483, 4803}

\bibitem[\protect\citeauthoryear{{Liao}, {Shafieloo}, {Keeley}  \&
  {Linder}}{{Liao} et~al.}{2019}]{2019ApJ...886L..23L}
{Liao} K.,  {Shafieloo} A.,  {Keeley} R.~E.,   {Linder} E.~V.,  2019, \mn@doi
  [\apjl] {10.3847/2041-8213/ab5308}, \href
  {https://ui.adsabs.harvard.edu/abs/2019ApJ...886L..23L} {886, L23}

\bibitem[\protect\citeauthoryear{{Liao}, {Shafieloo}, {Keeley}  \&
  {Linder}}{{Liao} et~al.}{2020}]{2020ApJ...895L..29L}
{Liao} K.,  {Shafieloo} A.,  {Keeley} R.~E.,   {Linder} E.~V.,  2020, \mn@doi
  [\apjl] {10.3847/2041-8213/ab8dbb}, \href
  {https://ui.adsabs.harvard.edu/abs/2020ApJ...895L..29L} {895, L29}

\bibitem[\protect\citeauthoryear{{Lukovi{\'c}}, {Haridasu}  \&
  {Vittorio}}{{Lukovi{\'c}} et~al.}{2020}]{2020MNRAS.491.2075L}
{Lukovi{\'c}} V.~V.,  {Haridasu} B.~S.,   {Vittorio} N.,  2020, \mn@doi
  [\mnras] {10.1093/mnras/stz3070}, \href
  {https://ui.adsabs.harvard.edu/abs/2020MNRAS.491.2075L} {491, 2075}

\bibitem[\protect\citeauthoryear{{Melia} \& {Yennapureddy}}{{Melia} \&
  {Yennapureddy}}{2018}]{2018JCAP...02..034M}
{Melia} F.,  {Yennapureddy} M.~K.,  2018, \mn@doi [\jcap]
  {10.1088/1475-7516/2018/02/034}, \href
  {https://ui.adsabs.harvard.edu/abs/2018JCAP...02..034M} {2018, 034}

\bibitem[\protect\citeauthoryear{{Millon}, {Galan}, {Courbin}, {Treu}, {Suyu}
  \& {et al.}}{{Millon} et~al.}{2020}]{2020A&A...639A.101M}
{Millon} M.,  {Galan} A.,  {Courbin} F.,  {Treu} T.,  {Suyu} S.~H.,   {et al.}
  2020, \mn@doi [\aap] {10.1051/0004-6361/201937351}, \href
  {https://ui.adsabs.harvard.edu/abs/2020A&A...639A.101M} {639, A101}

\bibitem[\protect\citeauthoryear{{Moresco}}{{Moresco}}{2015}]{2015MNRAS.450L..16M}
{Moresco} M.,  2015, \mn@doi [\mnras] {10.1093/mnrasl/slv037}, \href
  {https://ui.adsabs.harvard.edu/abs/2015MNRAS.450L..16M} {450, L16}

\bibitem[\protect\citeauthoryear{{Moresco}, {Cimatti}, {Jimenez}, {Pozzetti},
  {Zamorani}  \& {et al.}}{{Moresco} et~al.}{2012}]{2012JCAP...08..006M}
{Moresco} M.,  {Cimatti} A.,  {Jimenez} R.,  {Pozzetti} L.,  {Zamorani} G.,
  {et al.} 2012, \mn@doi [\jcap] {10.1088/1475-7516/2012/08/006}, \href
  {https://ui.adsabs.harvard.edu/abs/2012JCAP...08..006M} {2012, 006}

\bibitem[\protect\citeauthoryear{{Moresco}, {Pozzetti}, {Cimatti}, {Jimenez},
  {Maraston}  \& {et al.}}{{Moresco} et~al.}{2016}]{2016JCAP...05..014M}
{Moresco} M.,  {Pozzetti} L.,  {Cimatti} A.,  {Jimenez} R.,  {Maraston} C.,
  {et al.} 2016, \mn@doi [\jcap] {10.1088/1475-7516/2016/05/014}, \href
  {https://ui.adsabs.harvard.edu/abs/2016JCAP...05..014M} {2016, 014}

\bibitem[\protect\citeauthoryear{{Pandey}, {Karwal}  \& {Das}}{{Pandey}
  et~al.}{2020}]{2020JCAP...07..026P}
{Pandey} K.~L.,  {Karwal} T.,   {Das} S.,  2020, \mn@doi [\jcap]
  {10.1088/1475-7516/2020/07/026}, \href
  {https://ui.adsabs.harvard.edu/abs/2020JCAP...07..026P} {2020, 026}

\bibitem[\protect\citeauthoryear{Pedregosa, Varoquaux, Gramfort, Michel,
  Thirion  \& {et al.}}{Pedregosa et~al.}{2011}]{scikit-learn}
Pedregosa F.,  Varoquaux G.,  Gramfort A.,  Michel V.,  Thirion B.,   {et al.}
  2011, Journal of Machine Learning Research, 12, 2825

\bibitem[\protect\citeauthoryear{{Pedregosa}, {Varoquaux}, {Gramfort},
  {Michel}, {Thirion}  \& {et al.}}{{Pedregosa}
  et~al.}{2012}]{2012arXiv1201.0490P}
{Pedregosa} F.,  {Varoquaux} G.,  {Gramfort} A.,  {Michel} V.,  {Thirion} B.,
  {et al.} 2012, arXiv e-prints, \href
  {https://ui.adsabs.harvard.edu/abs/2012arXiv1201.0490P} {p. arXiv:1201.0490}

\bibitem[\protect\citeauthoryear{{Planck Collaboration}}{{Planck
  Collaboration}}{2020}]{2020A&A...641A...6P}
{Planck Collaboration} 2020, \mn@doi [\aap] {10.1051/0004-6361/201833910},
  \href {https://ui.adsabs.harvard.edu/abs/2020A&A...641A...6P} {641, A6}

\bibitem[\protect\citeauthoryear{{Planck Collaboration} et~al.,}{{Planck
  Collaboration} et~al.}{2017}]{2017A&A...607A..95P}
{Planck Collaboration} et~al., 2017, \mn@doi [\aap]
  {10.1051/0004-6361/201629504}, \href
  {https://ui.adsabs.harvard.edu/abs/2017A&A...607A..95P} {607, A95}

\bibitem[\protect\citeauthoryear{{Planck Collaboration} et~al.,}{{Planck
  Collaboration} et~al.}{2020a}]{2020A&A...641A...1P}
{Planck Collaboration} et~al., 2020a, \mn@doi [\aap]
  {10.1051/0004-6361/201833880}, \href
  {https://ui.adsabs.harvard.edu/abs/2020A&A...641A...1P} {641, A1}

\bibitem[\protect\citeauthoryear{{Planck Collaboration} et~al.,}{{Planck
  Collaboration} et~al.}{2020b}]{2020A&A...641A...7P}
{Planck Collaboration} et~al., 2020b, \mn@doi [\aap]
  {10.1051/0004-6361/201935201}, \href
  {https://ui.adsabs.harvard.edu/abs/2020A&A...641A...7P} {641, A7}

\bibitem[\protect\citeauthoryear{{Poulin}, {Smith}, {Karwal}  \&
  {Kamionkowski}}{{Poulin} et~al.}{2019}]{2019PhRvL.122v1301P}
{Poulin} V.,  {Smith} T.~L.,  {Karwal} T.,   {Kamionkowski} M.,  2019, \mn@doi
  [\prl] {10.1103/PhysRevLett.122.221301}, \href
  {https://ui.adsabs.harvard.edu/abs/2019PhRvL.122v1301P} {122, 221301}

\bibitem[\protect\citeauthoryear{{Rasmussen} \& {Williams}}{{Rasmussen} \&
  {Williams}}{2006}]{2006gpml.book.....R}
{Rasmussen} C.~E.,  {Williams} C. K.~I.,  2006, {Gaussian Processes for Machine
  Learning}

\bibitem[\protect\citeauthoryear{{Ratsimbazafy}, {Loubser}, {Crawford},
  {Cress}, {Bassett}  \& {et al.}}{{Ratsimbazafy}
  et~al.}{2017}]{2017MNRAS.467.3239R}
{Ratsimbazafy} A.~L.,  {Loubser} S.~I.,  {Crawford} S.~M.,  {Cress} C.~M.,
  {Bassett} B.~A.,   {et al.} 2017, \mn@doi [\mnras] {10.1093/mnras/stx301},
  \href {https://ui.adsabs.harvard.edu/abs/2017MNRAS.467.3239R} {467, 3239}

\bibitem[\protect\citeauthoryear{{Riess}}{{Riess}}{2020}]{2020NatRP...2...10R}
{Riess} A.~G.,  2020, \mn@doi [Nature Reviews Physics]
  {10.1038/s42254-019-0137-0}, \href
  {https://ui.adsabs.harvard.edu/abs/2020NatRP...2...10R} {2, 10}

\bibitem[\protect\citeauthoryear{{Riess}, {Casertano}, {Yuan}, {Macri}  \&
  {Scolnic}}{{Riess} et~al.}{2019}]{2019ApJ...876...85R}
{Riess} A.~G.,  {Casertano} S.,  {Yuan} W.,  {Macri} L.~M.,   {Scolnic} D.,
  2019, \mn@doi [\apj] {10.3847/1538-4357/ab1422}, \href
  {https://ui.adsabs.harvard.edu/abs/2019ApJ...876...85R} {876, 85}

\bibitem[\protect\citeauthoryear{{Riess} et~al.,}{{Riess}
  et~al.}{2022}]{2022ApJ...934L...7R}
{Riess} A.~G.,  et~al., 2022, \mn@doi [\apjl] {10.3847/2041-8213/ac5c5b}, \href
  {https://ui.adsabs.harvard.edu/abs/2022ApJ...934L...7R} {934, L7}

\bibitem[\protect\citeauthoryear{{Rigault} et~al.,}{{Rigault}
  et~al.}{2020}]{2020A&A...644A.176R}
{Rigault} M.,  et~al., 2020, \mn@doi [\aap] {10.1051/0004-6361/201730404},
  \href {https://ui.adsabs.harvard.edu/abs/2020A&A...644A.176R} {644, A176}

\bibitem[\protect\citeauthoryear{{Sapone}, {Nesseris}  \& {Bengaly}}{{Sapone}
  et~al.}{2021}]{2021PDU....3200814S}
{Sapone} D.,  {Nesseris} S.,   {Bengaly} C. A.~P.,  2021, \mn@doi [Physics of
  the Dark Universe] {10.1016/j.dark.2021.100814}, \href
  {https://ui.adsabs.harvard.edu/abs/2021PDU....3200814S} {32, 100814}

\bibitem[\protect\citeauthoryear{{Sch{\"o}neberg}, {Abell{\'a}n},
  {S{\'a}nchez}, {Witte}, {Poulin}  \& {Lesgourgues}}{{Sch{\"o}neberg}
  et~al.}{2022}]{2022PhR...984....1S}
{Sch{\"o}neberg} N.,  {Abell{\'a}n} G.~F.,  {S{\'a}nchez} A.~P.,  {Witte}
  S.~J.,  {Poulin} V.,   {Lesgourgues} J.,  2022, \mn@doi [\physrep]
  {10.1016/j.physrep.2022.07.001}, \href
  {https://ui.adsabs.harvard.edu/abs/2022PhR...984....1S} {984, 1}

\bibitem[\protect\citeauthoryear{Schulz, Speekenbrink  \& Krause}{Schulz
  et~al.}{2018}]{SCHULZ20181}
Schulz E.,  Speekenbrink M.,   Krause A.,  2018, \mn@doi [Journal of
  Mathematical Psychology] {https://doi.org/10.1016/j.jmp.2018.03.001}, 85, 1

\bibitem[\protect\citeauthoryear{{Scolnic}, {Jones}, {Rest}, {Pan}, {Chornock}
  \& {et al.}}{{Scolnic} et~al.}{2018}]{2018ApJ...859..101S}
{Scolnic} D.~M.,  {Jones} D.~O.,  {Rest} A.,  {Pan} Y.~C.,  {Chornock} R.,
  {et al.} 2018, \mn@doi [\apj] {10.3847/1538-4357/aab9bb}, \href
  {https://ui.adsabs.harvard.edu/abs/2018ApJ...859..101S} {859, 101}

\bibitem[\protect\citeauthoryear{{Seikel}, {Clarkson}  \& {Smith}}{{Seikel}
  et~al.}{2012}]{2012JCAP...06..036S}
{Seikel} M.,  {Clarkson} C.,   {Smith} M.,  2012, \mn@doi [\jcap]
  {10.1088/1475-7516/2012/06/036}, \href
  {https://ui.adsabs.harvard.edu/abs/2012JCAP...06..036S} {2012, 036}

\bibitem[\protect\citeauthoryear{{Sekiguchi} \& {Takahashi}}{{Sekiguchi} \&
  {Takahashi}}{2021}]{2021PhRvD.103h3507S}
{Sekiguchi} T.,  {Takahashi} T.,  2021, \mn@doi [\prd]
  {10.1103/PhysRevD.103.083507}, \href
  {https://ui.adsabs.harvard.edu/abs/2021PhRvD.103h3507S} {103, 083507}

\bibitem[\protect\citeauthoryear{{Shah}, {Lemos}  \& {Lahav}}{{Shah}
  et~al.}{2021}]{2021A&ARv..29....9S}
{Shah} P.,  {Lemos} P.,   {Lahav} O.,  2021, \mn@doi [\aapr]
  {10.1007/s00159-021-00137-4}, \href
  {https://ui.adsabs.harvard.edu/abs/2021A&ARv..29....9S} {29, 9}

\bibitem[\protect\citeauthoryear{{Shanks}, {Hogarth}  \& {Metcalfe}}{{Shanks}
  et~al.}{2019}]{2019MNRAS.484L..64S}
{Shanks} T.,  {Hogarth} L.~M.,   {Metcalfe} N.,  2019, \mn@doi [\mnras]
  {10.1093/mnrasl/sly239}, \href
  {https://ui.adsabs.harvard.edu/abs/2019MNRAS.484L..64S} {484, L64}

\bibitem[\protect\citeauthoryear{{Simon}, {Verde}  \& {Jimenez}}{{Simon}
  et~al.}{2005}]{2005PhRvD..71l3001S}
{Simon} J.,  {Verde} L.,   {Jimenez} R.,  2005, \mn@doi [\prd]
  {10.1103/PhysRevD.71.123001}, \href
  {https://ui.adsabs.harvard.edu/abs/2005PhRvD..71l3001S} {71, 123001}

\bibitem[\protect\citeauthoryear{{Spergel}, {Gehrels}, {Breckinridge},
  {Donahue}, {Dressler}  \& {et al.}}{{Spergel}
  et~al.}{2013}]{2013arXiv1305.5422S}
{Spergel} D.,  {Gehrels} N.,  {Breckinridge} J.,  {Donahue} M.,  {Dressler} A.,
    {et al.} 2013, arXiv e-prints, \href
  {https://ui.adsabs.harvard.edu/abs/2013arXiv1305.5422S} {p. arXiv:1305.5422}

\bibitem[\protect\citeauthoryear{{Suyu}, {Chang}, {Courbin}  \&
  {Okumura}}{{Suyu} et~al.}{2018}]{2018SSRv..214...91S}
{Suyu} S.~H.,  {Chang} T.-C.,  {Courbin} F.,   {Okumura} T.,  2018, \mn@doi
  [\ssr] {10.1007/s11214-018-0524-3}, \href
  {https://ui.adsabs.harvard.edu/abs/2018SSRv..214...91S} {214, 91}

\bibitem[\protect\citeauthoryear{{Vagnozzi}}{{Vagnozzi}}{2021}]{2021PhRvD.104f3524V}
{Vagnozzi} S.,  2021, \mn@doi [\prd] {10.1103/PhysRevD.104.063524}, \href
  {https://ui.adsabs.harvard.edu/abs/2021PhRvD.104f3524V} {104, 063524}

\bibitem[\protect\citeauthoryear{{Verde}, {Treu}  \& {Riess}}{{Verde}
  et~al.}{2019}]{2019NatAs...3..891V}
{Verde} L.,  {Treu} T.,   {Riess} A.~G.,  2019, \mn@doi [Nature Astronomy]
  {10.1038/s41550-019-0902-0}, \href
  {https://ui.adsabs.harvard.edu/abs/2019NatAs...3..891V} {3, 891}

\bibitem[\protect\citeauthoryear{{Wang} \& {Dai}}{{Wang} \&
  {Dai}}{2013}]{Wang2013}
{Wang} F.~Y.,  {Dai} Z.~G.,  2013, \mn@doi [\mnras] {10.1093/mnras/stt652},
  \href {https://ui.adsabs.harvard.edu/abs/2013MNRAS.432.3025W} {432, 3025}

\bibitem[\protect\citeauthoryear{{Wang}, {Hu}, {Zhang}  \& {Dai}}{{Wang}
  et~al.}{2022}]{2022ApJ...924...97W}
{Wang} F.~Y.,  {Hu} J.~P.,  {Zhang} G.~Q.,   {Dai} Z.~G.,  2022, \mn@doi [\apj]
  {10.3847/1538-4357/ac3755}, \href
  {https://ui.adsabs.harvard.edu/abs/2022ApJ...924...97W} {924, 97}

\bibitem[\protect\citeauthoryear{{Wong}, {Suyu}, {Chen}, {Rusu}, {Millon}  \&
  {et al.}}{{Wong} et~al.}{2020}]{2020MNRAS.498.1420W}
{Wong} K.~C.,  {Suyu} S.~H.,  {Chen} G. C.~F.,  {Rusu} C.~E.,  {Millon} M.,
  {et al.} 2020, \mn@doi [\mnras] {10.1093/mnras/stz3094}, \href
  {https://ui.adsabs.harvard.edu/abs/2020MNRAS.498.1420W} {498, 1420}

\bibitem[\protect\citeauthoryear{{Wu}, {Zhang}  \& {Wang}}{{Wu}
  et~al.}{2022}]{2022MNRAS.515L...1W}
{Wu} Q.,  {Zhang} G.-Q.,   {Wang} F.-Y.,  2022, \mn@doi [\mnras]
  {10.1093/mnrasl/slac022}, \href
  {https://ui.adsabs.harvard.edu/abs/2022MNRAS.515L...1W} {515, L1}

\bibitem[\protect\citeauthoryear{{Yu}, {Ratra}  \& {Wang}}{{Yu}
  et~al.}{2018}]{2018ApJ...856....3Y}
{Yu} H.,  {Ratra} B.,   {Wang} F.-Y.,  2018, \mn@doi [\apj]
  {10.3847/1538-4357/aab0a2}, \href
  {https://ui.adsabs.harvard.edu/abs/2018ApJ...856....3Y} {856, 3}

\bibitem[\protect\citeauthoryear{{Zhang}, {Zhang}, {Yuan}, {Liu}, {Zhang}  \&
  {et al.}}{{Zhang} et~al.}{2014}]{2014RAA....14.1221Z}
{Zhang} C.,  {Zhang} H.,  {Yuan} S.,  {Liu} S.,  {Zhang} T.-J.,   {et al.}
  2014, \mn@doi [Research in Astronomy and Astrophysics]
  {10.1088/1674-4527/14/10/002}, \href
  {https://ui.adsabs.harvard.edu/abs/2014RAA....14.1221Z} {14, 1221}

\bibitem[\protect\citeauthoryear{{de Jaeger}, {Galbany}, {Riess}, {Stahl},
  {Shappee}, {Filippenko}  \& {Zheng}}{{de Jaeger}
  et~al.}{2022}]{2022MNRAS.514.4620D}
{de Jaeger} T.,  {Galbany} L.,  {Riess} A.~G.,  {Stahl} B.~E.,  {Shappee}
  B.~J.,  {Filippenko} A.~V.,   {Zheng} W.,  2022, \mn@doi [\mnras]
  {10.1093/mnras/stac1661}, \href
  {https://ui.adsabs.harvard.edu/abs/2022MNRAS.514.4620D} {514, 4620}

\makeatother
\end{thebibliography}
\bsp	

\label{lastpage}
\end{document}